\documentclass[manuscript]{aastex} \usepackage{amsmath}

\shorttitle{3D view of THMF in the photosphere} \shortauthors{Ishikawa,
Tsuneta, \& Jan Jur\v{c}\'{a}k}

\begin{document}
\title{3D view of transient horizontal magnetic fields in the photosphere}
\author{Ryohko Ishikawa \altaffilmark{1,2} \and Saku Tsuneta\altaffilmark{2} \and Jan Jur\v{c}\'{a}k\altaffilmark{3}}
\email{ryoko.ishikawa@nao.ac.jp}

\altaffiltext{1}{Department of Astronomy, University of Tokyo, Hongo,
Bunkyo-ku, Tokyo 113-0033, Japan} \altaffiltext{2}{National Astronomical
Observatory of Japan, 2-21-1 Osawa, Mitaka, Tokyo 181-8588, Japan}
\altaffiltext{3}{Astronomical Institute of the Academy of Sciences, Fri\v{c}ova
298, 25165 Ond\v{r}ejov, Czech Republic}

\begin{abstract}
We infer the 3D magnetic structure of a \emph{transient
horizontal magnetic field} (THMF) during its evolution through the photosphere
using SIRGAUS inversion code.  
The SIRGAUS code is a modified version of SIR 
(Stokes Inversion based on Response function), and
allows for retrieval of information on the magnetic and thermodynamic parameters of the flux 
tube embedded in the atmosphere from the observed Stokes profiles. 
Spectro-polarimetric observations of the quiet Sun at the disk center were performed with the Solar Optical Telescope (SOT) on board \emph{Hinode} with \ion{Fe}{1} 630.2~nm lines. 
Using repetitive scans with a cadence of 130~s, we first detect
the horizontal field that appears inside a granule, near its edge. 
On the second scan, vertical fields with
positive and negative polarities appear at both ends of the horizontal field.
Then, the horizontal field disappears leaving the bipolar
vertical magnetic fields. The results from the inversion of the Stokes spectra clearly point to the
existence of a flux tube with magnetic field strength of $\sim400$~G
rising through the line forming layer of the  \ion{Fe}{1} 630.2~nm lines. The flux tube is
located at around $\log\tau_{500} \sim0$ at $\Delta t$=0~s and  
around $\log\tau_{500} \sim-1.7$ at $\Delta t$=130~s.
At $\Delta t$=260~s the horizontal part is already
above the line forming region of the analyzed lines. 
The observed Doppler velocity is maximally 3~km~s$^{-1}$, 
consistent with the upward motion of the structure as retrieved from the SIRGAUS
code.
The vertical size of the tube
is smaller than the thickness of the line forming layer. The THMF has a clear
$\Omega$-shaped-loop structure with the apex located near the edge of a
granular cell. The magnetic flux
carried by this THMF is estimated to be $3.1\times10^{17}$ Mx. 
\end{abstract}

\section{Introduction}
The SOT spectropolarimeter
\citep[SP,][]{Tsuneta2008SoPh,Suematsu2008SoPh,Ichimoto2008SoPh,Shimizu2008SoPh} on
board \emph{Hinode} \citep{Kosugi2007} revealed that there are ubiquitous
horizontally inclined magnetic fields with strengths of a few hundred G in the
internetwork regions \citep{Orozco2007,Lites2007}. \citet{Centeno2007} and
\citet{Ishikawa2008} reported the temporal evolution of such horizontal
magnetic fields both in the quiet Sun and in plage regions. The evolution commonly starts
with the appearance of linear polarization signals (horizontal fields) inside
granular cells, followed by the appearance of the opposite-polarity circular
polarization signals (vertical fields) at both ends of the horizontal
component. The time evolution indicates that the horizontally inclined magnetic
fields are part of a small magnetic loop. The sizes of the horizontal magnetic
fields are smaller than the size of the granule where they appear.
Statistical analyses of a large number of the horizontal fields indicate that
the life time of the horizontal magnetic fields ranges from 1 to 10 minutes
and the mean lifetime is about 4 minutes
\citep{Ishikawa2009,Ishikawa_Hinode2}. Their occurrence rate is high, and more than
approximately 10\% of the granules have embedded horizontal magnetic fields
\citep{Ishikawa2008}. The movie of linear polarization signals \citep{Ishikawa2009} shows that these
horizontal magnetic fields are highly frequent and transient, and the term
''transient horizontal magnetic fields'' (THMFs) is used for them. Earlier reports
of flux emergence in the quiet Sun
\citep{Lites1996,DePontieu2002,MartinezGonzalez2007} appear to be examples of
THMFs.

The properties of the THMFs described above are the same in the quiet Sun and the plage region.
There are granular-sized horizontal fields in the polar regions, and these magnetic fields may be the same as THMFs \citep{Tsuneta2008ApJ,Itoh2009}.
However, there seem to be a few distinct patterns in
their disappearance. For example, \citet{Ishikawa2008} report an example of a 
THMF disappearing in the intergranular lane, while \citet{Centeno2007} show a
THMF which disappears inside a bright granule. This is confirmed by
\citet{Ishikawa_Hinode2} who report that normalized continuum intensities
where THMFs appear are above $\sim$1.0, corresponding to the bright granular
regions. However, the normalized continuum intensities where THMFs disappear
are both higher and lower than $\sim$1.0, i.e. THMFs do not necessarily reach
the intergranular lanes.
There also appears to be two types of THMFs; with or without apparent
bipolar vertical fields after their disappearance. This variety of THMF
behavior suggests a few different scenarios with regard to their disappearance:
(1) THMFs somehow submerge with the downward convective motion; (2) They
become fragmentary due to the convective flow, and their Stokes signals are
below the SP detection level; (3) THMFs go through the line formation layer
of the \ion{Fe}{1}~630.2~nm lines and reach higher atmospheric layers \citep{MartinezGonzalez2009}.  

Given the ubiquity of THMFs in the solar photosphere, the question of how they appear and disappear in such a short time becomes one of the important issues in solar magnetohydrodynamics (MHD). A series of sophisticated MHD simulations
\citep[e.g.,][]{Vogler2007, Abbett2007, Schussler2008, Isobe2008, Steiner2008}
have been carried out with varieties of initial and boundary conditions. A detailed
comparison of the observed properties of THMFs with the numerical simulation
results has just begun. It is thus desirable to clarify the structure and
evolution of the enigmatic small-scale magnetic fields in the photosphere, and
possibly in the chromosphere.

Spectropolarimetric data obtained with the SOT have been analyzed thus far mainly
with inversion codes based on a Milne-Eddington atmosphere
\citep[e.g.,][]{Lites2007, Orozco2007, Ishikawa2009}. Milne-Eddington inversion
provides us with various physical parameters (magnetic field vector and Doppler
velocity, etc.), which have to be constant throughout the atmosphere (except for
the source function), and synthesize only symmetric Stokes profiles. However,
the observed Stokes profiles of the horizontal magnetic fields usually have  
area-asymmetries. Such asymmetric profiles suggest the presence of velocity and magnetic
field gradients along the line of sight (LOS) \citep[e.g.,][]{Solanki1988,Sanchez1992}. Thus using the Milne-Eddington
inversion, physically important information about the small-scale
horizontal magnetic fields may be lost. 

We are using an inversion code based on the SIR code
\citep[Stokes Inversion based on Response
function,][]{RuizCobo1992,RuizCobo1994,DelToroIniesta1996,DelToroIniest2003}.
This code allows for the change of atmospheric parameters along the LOS and thus
fits the observed asymmetric Stokes profiles. \citet{MartinezGonzalez2007} used
the SIR inversion code to analyze a snapshot of the magnetic loops using the
spectropolarimetric data of 1.56 $\mu$m taken with the Tenerife Infrared
Polarimeter. However, the used settings of the inversion assumed the plasma
parameters to be constant with height in the atmosphere, except in magnetic
field inclination and temperature.
Recently, \citet{Gomory2009} studied the temporal evolution of a similar event applying SIR inversions.
However, they kept all magnetic parameters constant with height.
Thus, these analyses resemble a Milne-Eddington inversion with a realistic temperature 
distribution. We fully exploit the inversion code without such constraints to analyze the 
observed Stokes profiles asymmetries and with a better spatial resolution. This allows 
us to follow the temporal evolution of the stratified atmosphere.

In this paper, we study a THMF event whose properties are similar to those reported by
\citet{Centeno2007}. We carry out the full Stokes inversion with the SIRGAUS code 
\citep{BellotRubio2003}, which is a modified version of SIR. The SIRGAUS code treats 
explicitly a magnetic flux tube embedded in the atmosphere with a Gaussian function, and is
expected to be more suitable to obtain vertical (LOS) structure of the THMF. Indeed, we 
successfully identify an isolated flux tube in the photosphere, and track the emergence of
 the $\Omega$-shaped flux tube through the photosphere. We describe statistical significance 
 of our result in detail. With the clear identification of the upward moving flux tube, we are able
 to obtain various physical parameters that characterize the flux tube. Such information on the
 flux tube becomes available for the first time with the use of the SIRGAUS code. 
 We finally discuss the magnetohydrodynamic properties of the THMF.

\section{Observations and wavelength calibration}
\label{wavelengthcalib} A quiet solar region of $4\arcsec \times 76\arcsec$
located near the Sun center was observed with a cadence of 130 s using the
SOT/SP on March 10, 2007. We obtained the Stokes spectra of the \ion{Fe}{1}~630.2 nm
lines, whose line formation region is between $\log\tau_{500}\sim$ 0 and ~$-2.5$ 
\citep{CabreraSolana2005}.  
Each slit position has an exposure time of 4.8~s, and the scanning step is 0\farcs15. 
The noise level in the Stokes $Q$ and $U$ signal is
$1.2\times10^{-3}I_{c}$ and that in the Stokes $V$ is $1.1\times10^{-3}I_{c}$,
where $I_{c}$ is the continuum intensity. The dark current, the flat field
corrections, and the polarimetric calibration were performed using
``sp\_prep" software.

\begin{figure}
\epsscale{.8}
\plotone{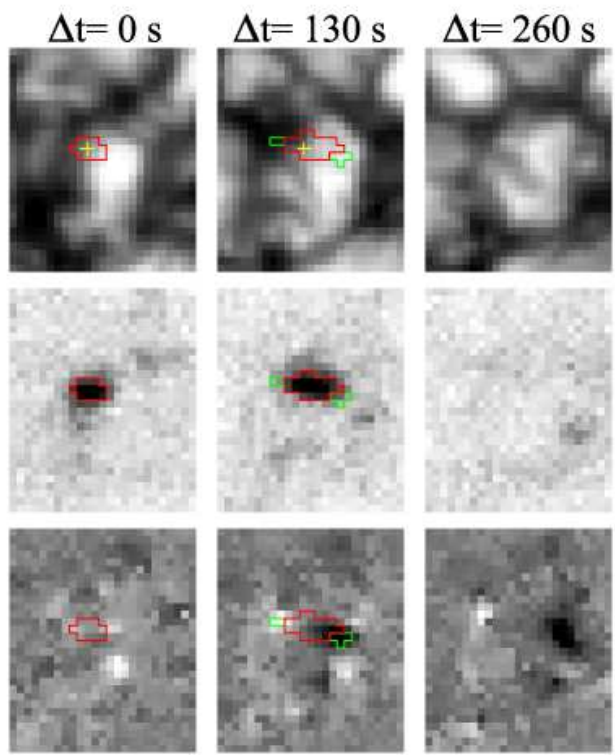}
 \caption{Temporal evolution of the continuum $I_{c}$ ($top$) and the integrated
linear/circular polarization signal ($middle/bottom$).  
The integration range for the linear polarization signal is from
$-21.4$~pm to $21.4$~pm from the line center of the Stokes $I$ spectra
at each pixel. The integration for the circular polarization is carried
out from $-21.4$~pm to $-4.28$~pm and from 4.28~pm to 21.4~pm, taking
the absolute values of the Stokes V spectra. The sign for the obtained
circular polarization signal is determined to be the sign of the peak
Stokes $V$ profile at the blue wing.
The time cadence is 130 s. The pixels to be
analyzed in this paper are shown with red and green contours (see text).}
\label{eventfig}
\end{figure}

We extensively analyze the event shown in Figure~\ref{eventfig}.
The horizontal magnetic field appears at the edge of a granule without any
vertical magnetic fields ($\Delta t=0$ s). At $\Delta t=130$ s (second frame),
the size of the horizontal magnetic field becomes larger and the bipolar
vertical fields appear at both ends of the horizontal field. At the third frame
($\Delta t=260$ s), the horizontal field disappears, leaving an isolated pair
of bipolar vertical fields. The separation of these two patches increases
with time. The lifetime of the horizontal magnetic field is about 4 min, and
this event has properties in common with those of the THMFs as reported by
 \citet{Ishikawa2008}.

We analyze only the pixels with the Stokes $Q$ or $U$ amplitude 10 times larger
than their noise level. Such pixels
are shown with the red contours in Figure~\ref{eventfig}. Pixels shown with the
green contours are dominated by the Stokes $V$ signal, and correspond to the
footpoints of the THMF. These pixels are also analyzed.

The Doppler velocity information is crucial for understanding the dynamics of the
flux tube. For this purpose, accurate in-orbit calibration of the SOT/SP
wavelength scale and offset is done to obtain the absolute velocity scale. We
here calibrate our data using the solar flux atlas \citep{Beckers1976}. The
atlas contains a high resolution spectra of the solar irradiance from 380~nm to
700~nm and has the wavelength information.

\begin{figure}
\plotone{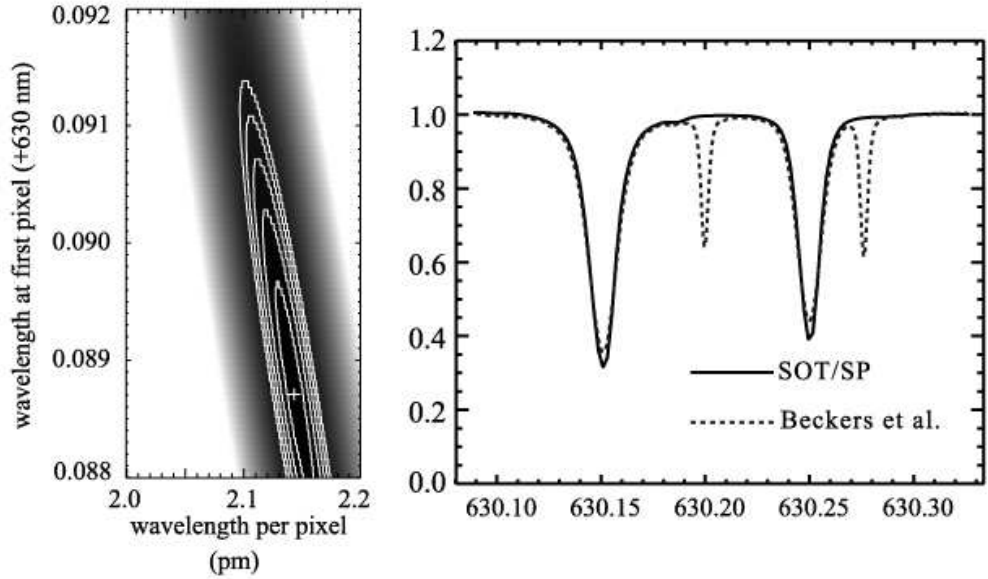}
 \caption{$Left:$ Correlation coefficient for the combination of the
SOT/SP wavelength per pixel and the wavelength at the first pixel (offset).
The contours show the correlation coefficient of 0.929, 0.931,
0.933, 0.935, and 0.937. The cross represents the position with the maximum
correlation coefficient of 0.938, in which the wavelength per pixel and the
wavelength at the first pixel are 2.14~pm pix$^{-1}$ and 630.089~nm. $Right:$
The SOT/SP mean Stokes $I$ profile (see text), which is wavelength-calibrated
compared to the solar atlas
\citep{Beckers1976}.  Note that these profiles are normalized to the continuum
intensity.} \label{wavecal}
\end{figure}

First, we obtain the mean Stokes $I$ profile of the quiet Sun by averaging the
Stokes $I$ profiles of pixels without the polarization signal. Second, we
compare the mean Stokes $I$ profile with the solar flux atlas and derive the
wavelength sampling of 2.14~pm and the wavelength value at the start pixel of
630.089~nm. These two parameters give us the maximum correlation coefficient
between the mean Stokes $I$ profile obtained with the SOT/SP and the solar flux
atlas (see Fig. \ref{wavecal}). However, the wavelength shift of the flux atlas is
not corrected for the gravitational redshift and convective blueshift
\citep{Allende1998}. It turns out that the
two \ion{Fe}{1}~lines have the redshifts of 0.88~pm and 0.63~pm (420~m~s$^{-1}$ and
300~m~s$^{-1}$) with respect to the laboratory centers, respectively. These
different Doppler shifts come from the difference in the convective blueshifts
between the two lines. 
Thus, we define the mean value of the two redshifts for the two \ion{Fe}{1}~lines to be 360~m~s$^{-1}$ as the zero point of the velocity scale. With this process, we have corrected the effect of the gravitational redshift and the convective blueshift. The Doppler velocities in this paper are now defined such that if there is no motion on the Sun, 
this will give the zero velocity.

\section{SIRGAUS Inversion}
\label{inversion}
We perform the inversion of the observed Stokes profiles using the SIRGAUS code \citep{BellotRubio2003}. This inversion code synthesizes the Stokes profiles emerging from a model atmosphere and then compares them to the observed ones. Using the least-square  Levenberg-Marquardt algorithm, the initial model is modified until the difference between the observed and the synthetic profiles is minimized. The model consists of a simple atmosphere that represents the $background$ and a Gaussian-shaped bump superposed on some of the atmospheric parameters representing an embedded flux tube. Note that we are considering a one-component model atmosphere. The THMF analyzed here is spread out over several pixels. Therefore, we assume that the magnetic filling factor, $f$, is one.

Recent Milne-Eddington inversions of THMFs have provided an average filling factor of about 0.2 \citep{Ishikawa2009}. The observed magnetic structures we want to analyze cover more than a spatial resolution element of the SP data. Thus, if we neglect stray-light contamination effects like those due to diffraction \citep{Orozco2007PASJ} or scattered light, we would expect $f$ to be around unity. The obtained small filling factors might be just a net effect of the used technique that cannot account for a flux tube fully covering
the resolution element, but only partially occupying the line forming region of the \ion{Fe}{1} lines at 630.2~nm.

Thus far, this code has not been widely used. Examples of its application to spectropolarimetric data can be found in, e.g., \citet{Jurcak2007} where they analyze Stokes profiles observed in Sunspot penumbrae. As explained before, in SIRGAUS the distribution of the atmospheric parameters, $x$, consists of a background atmosphere and of a Gaussian bump, and can be written as
\begin{align}
x(\eta)= x^{b}(\eta)+\Delta x \exp
\left(-\frac{(\eta-G_{pos})^{2}}{G_{width}^{2}}\right),
\end{align}
where $x^{b}$ stands for the background plasma parameters; $\Delta x$ represents the amplitude of the Gaussian bump; $\eta$ is the logarithm of the optical depth, $\tau_{500}$;  and $G_{width}$ and $G_{pos}$ are the width and the position of the Gaussian bump in $\eta$ units.  The value of the amplitude, $\Delta x$,  is different for each of the physical quantities. The Gaussian bump position and width are the same for all the physical quantities.

The SIRGAUS code tries to find the best model atmosphere that reproduces the observed profiles in a  space of 14 free parameters: The temperature, $T$, of the background atmosphere is allowed to change at three nodes\footnote{See e.g., \citet{RuizCobo1992}}. The other free parameters, $x^{b}$, characterizing the background atmosphere, say the microturbulence, $v_{mic}^{b}$, the LOS velocity, $v_{LOS}^{b}$, and the field strength, $B^{b}$,  its inclination, $\gamma^{b}$, and azimuth, $\phi^{b}$, are assumed to be constant with height. The rest of the free parameters are the amplitudes of the bumps, $\Delta v_{LOS}$, $\Delta B$, $\Delta \gamma$, and $\Delta \phi$, and their width, $G_{width}$, and position, $G_{pos}$.  Hereafter and for simplicity, since the Gaussian bump describes the plasma parameters of a flux tube embedded in the line forming region of the observed spectral lines, the Gaussian bump will be referred to as flux tube.

Throughout this paper, we are displaying the resulting distributions of
plasma parameters as a function of the logarithm of the continuum optical depth
at 500~nm; $\log\tau_{500}$, hereafter referred to as $\log\tau$.
However, the optical depths will be converted to kilometers to specific purposes (see e.g., Sect.~\ref{Magflux}). 
To this end we solve the following equation: $\tau= \int_0^z \kappa \rho dz$.
With the boundary condition $\tau=1$ at $z=0$~km,
$\kappa$ is the continuum absorption
coefficient per unit mass and $\rho$ the gas density. 
The absorption coefficient depends on different plasma parameters and on the chemical composition.
The results from the SIRGAUS inversion are used to evaluate the previous equation at each pixel.
For its integration we use the program geometrical.x, provided with the SIR code.
The same program also provides us with the values of the gas density that will be used later in this paper (see Sect.~\ref{forcebalance}).

\section{Emergence of horizontal magnetic fields}
\label{emerging} 
\begin{figure}
\plotone{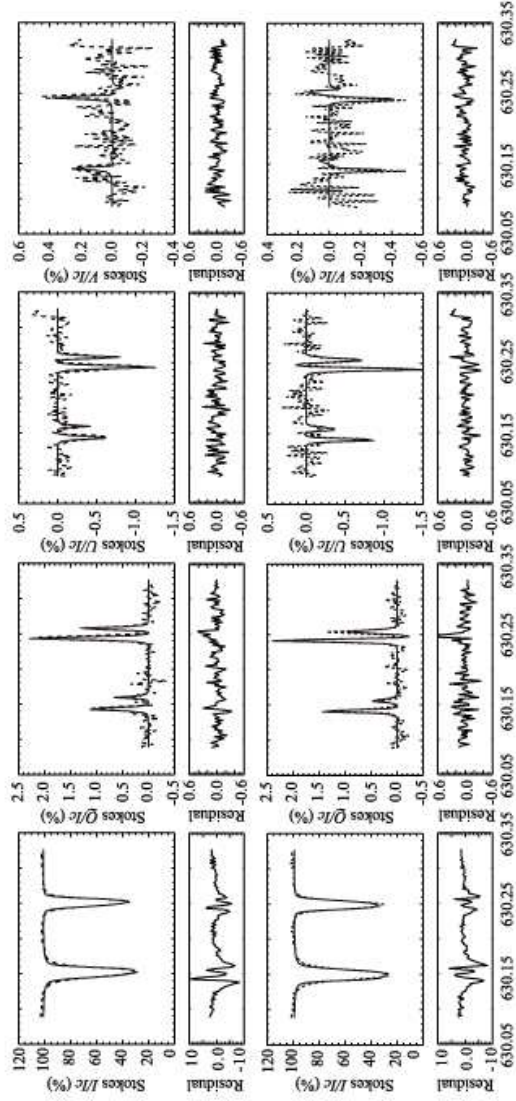}
  \caption{Observed (dashed) and the synthetic (solid) Stokes
profiles for pixel\#1 from the first scan ($top$) and pixel\#2 from the second scan
($bottom$). At the bottom of each panel, the difference between the observed and
the synthetic profiles is shown in percent. } \label{prof}
\end{figure}

In order to clearly show the evolution of the horizontal
magnetic field between the two scans, we select two pixels, one in each scan,
and present their detailed analysis. The selected pixels are marked by yellow
crosses in Figure~\ref{eventfig}. 
The corresponding observed Stokes profiles are shown in Figure~\ref{prof}. 
To invert the Stokes profiles of all pixels we used several initial atmospheric models. 
For the pixel taken from the first scan (hereafter pixel\#1), all the initial models converged to a
solution where the flux tube is located low in the atmosphere. On the other
hand, the solutions for the pixel from the second scan (pixel\#2) show that the
position of the flux tube is located higher in the photosphere.

To show the reliability of the obtained positions and widths of the flux tube, we
proceeded as follows: First we fixed all the plasma parameters obtained by the SIRGAUS
inversion. Then we evaluated a merit  function for a wide range of
$G_{width}$ and $G_{pos}$ values. For this purpose, we introduced a $\chi^{2}$
function that does not take into account the contribution of the Stokes $I$ profile:
\begin{eqnarray}
\chi^{2} \equiv \sum_{k=1}^{3}\sum_{l=1}^{n}
\frac{[I_{k}^{obs}(\lambda_{l})-I_{k}^{syn}(\lambda_{l})]^{2}}{\sigma_{o}^{2}},
\label{eq:chi}
\end{eqnarray}
where $n$ is the number of the wavelength points; $I_{k}$ with $k=1,2,3$
indicates the Stokes $Q$, $U$, and $V$profiles; $I_{k}^{obs}$ and $I_{k}^{syn}$
stands for the observed and the synthetic profiles, respectively; and
$\sigma_{o}$ is the standard deviation of the Stokes $Q$, $U$, and $V$ profiles
estimated using pixels without any apparent polarization signal. It
represents the Poisson noise of incident photons and the value is
$\sim1.1\times10^{-3}I_{c}$ (See Sect. \ref{wavelengthcalib}).

Hereafter, we use the $\chi^{2}$ function as defined in Eqn. (\ref{eq:chi}).
The SIRGAUS code minimizes a $\chi^{2}$ that takes into account all four
Stokes profiles. The synthetic Stokes profiles shown in Figure~\ref{prof} are
example fits to the observarions. As can be seen in the lower panels (residuals), the
fits are very good for the $Q$, $U$, and $V$ profiles and reasonable for the $I$ profile. 
The largest discrepancies
are obtained in the wings of the Stokes $I$ profile with a maximum of about 10\%.
The origin of the residual error in the wings of Stokes $I$ may be due to not taking
into account the hydrogen collisional broadening \citep{Bellot2009}.
However, it should be noted that SIRGAUS retrieves the information about
the atmospheric structure along the LOS mainly from the asymmetries of the
Stokes $Q$, $U$, and $V$ profiles. For this reason we use a 
$\chi^{2}$ based only on the Stokes $Q$, $U$, and $V$ profiles.
This $\chi^{2}$ definition will also help in clarifying the properties of the flux tube.

\begin{figure}
\plotone{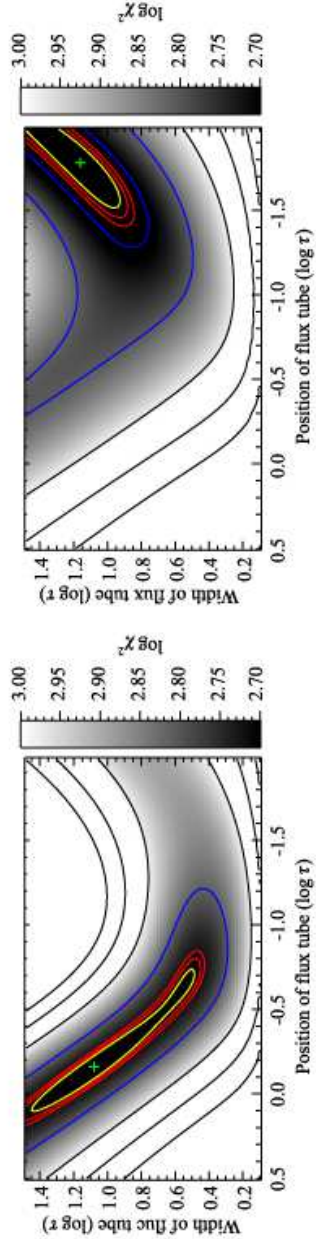} \caption{$\chi^{2}$ maps in the parameter space of the
position and width of the flux tube for pixel\#1 ($\Delta t=0$ s, $left$)
and pixel\#2 ($\Delta t=130$ s, $right$). The cross shows the position where the
$\chi^{2}$ is minimum. The yellow line indicates
$\Delta \chi^{2}=13.7$ ($1\sigma$) and the red ones $\Delta \chi^{2}=21.3$
($2\sigma$) and 29.8 ($3\sigma$).  Blue contours indicate the reduced
$\chi^{2}$ values of 1.5 and 2.0, while black ones correspond to
3.0, 4.0, and 5.0. We define the reduced $\chi^{2}$ as
 $\chi^{2}/(3n-p)$, where $n$ is the number of the wavelength
points, and $p$ the number of free parameters. }
 \label{chi}
\end{figure}

\begin{figure}
\plotone{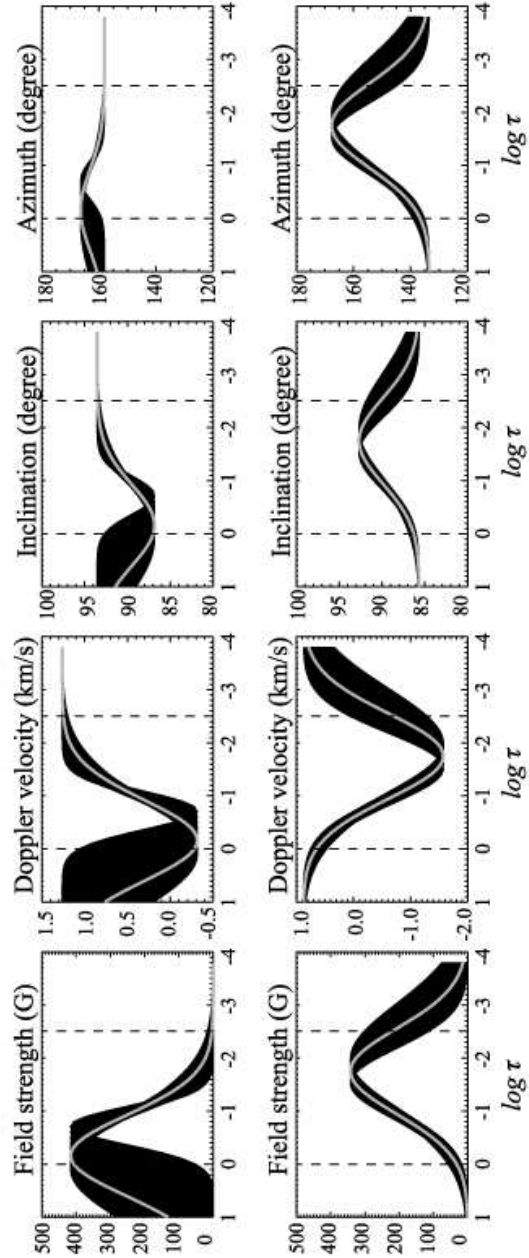} \caption{Various physical parameters as a function of the optical depth obtained for pixel\#1 ($top$) and pixel\#2 ($bottom$). 
The gray lines show the parameters with the smallest $\chi^{2}$ value.
The black areas are created by plotting the distributions of plasma parameters within the $1\sigma$ confidence regions. 
The dashed lines enclose the line forming region of the  \ion{Fe}{1}~630.2~nm lines.}
  \label{atmos}
\end{figure}

Figure~\ref{chi} shows the dependence of the $\chi^{2}$ values on the position and
width of the flux tube for pixel\#1 and pixel\#2 (left and right panels, respectively). 
The minimum values ($\chi^{2}_{min}$) are marked by the
crosses and correspond to the synthetic profiles shown in Figure~\ref{prof} and to the models
shown by the gray lines in Figure~\ref{atmos}. 
To construct Figure~\ref{chi}, we fix 12 of the 14 free parameters and evaluate the 
$\chi^{2}$ value using Eqn. (\ref{eq:chi}) and varying the $G_{pos}$ and $G_{width}$. 
The solid curves in Figure~\ref{chi} are defined by constant values of
$\Delta\chi^{2}\equiv \chi^{2}-\chi^{2}_{min}$. The region outlined by
the $\Delta\chi^{2}$ curves defines the level of confidence: for instance, there is a
68.3\% ($1\sigma$) chance for the true parameters of the position and
width of the flux tube to fall within the region defined by the yellow contour
($\Delta \chi^{2}=13.7$). Likewise, there is $95.4\%$ ($2\sigma$) and
$99.7\%$ ($3\sigma$) chance that the true parameters fall inside the red
contours with $\Delta \chi^{2}=21.30$ and 29.8, respectively. 
As explained before to make Figure~\ref{chi}, we have varied 2 parameters  (the positions and widths of the flux tube) and remaining 12 parameters are fixed.
In this case, $\Delta \chi^{2}$ is distributed as a chi-square
distribution with $14-2=12$ degrees of freedom 
\citep[TheoremD in][]{Numerical}. The $\Delta \chi^{2}=$13.7, 21.3, and 29.8 used above
come from the chi-square distribution function for 12 degrees of freedom and for the confidence level of 68.27$\%$, 95.45$\%$, and 99.73$\%$, respectively.

In pixel\#1 ($\Delta t=0$~s), the $\chi^{2}$ is minimum when the
position of the flux tube is about $\log\tau \sim-0.15$ and its width is about 1.1.
However, these parameters are not determined uniquely since the $1\sigma$
confidence level is elongated. The positions of the flux tube extend
from $\log\tau \sim0.1$ to $-0.7$ with $G_{width}$ changing accordingly from 1.4
to 0.5. A thick flux tube located deeper in the photosphere and a thin flux
tube located higher in the atmosphere provide equally good fits to the observed spectra. 
From this results we can ascertain that  
the center of the flux tube is not located higher than $\log\tau
\sim-0.7$. This optical depth corresponds to a geometrical height of 91~km.

The $1\sigma$ confidence level is less elongated for pixel\#2 ($\Delta
t=130$~s), but again, it is difficult to specify the exact width and position of the flux
tube. The minimum $\chi^{2}$ is obtained when the position of the flux tube is about
$\log\tau \sim-1.8$ and its width about 1.1. The position of the flux tube is
definitely higher than $\log\tau \sim-1.5$, i.e., 213~km.
In conclusion, the
$\chi^{2}$ maps clearly show with sufficient statistical significance that the magnetic flux 
tube is located higher in the atmosphere at $\Delta t=130$~s
than at $\Delta t=0$~s. Therefore, the flux tube moves upwards with
time.

The distributions of plasma parameters as a function of the optical depth 
are shown in Figure~\ref{atmos}. The gray line stands for the atmospheric 
model at $\chi^{2}_{min}$.  The black areas represent the variation 
within the 1$\sigma$ confidence region.
We first notice that the flux tube does not occupy the whole line forming region.
For pixel\#1, the center of the flux tube is around $\log\tau \sim -0.5$ 
and we are observing the upper boundary of the flux tube. On the other hands, for pixel\#2,
the flux tube is located at around $\log\tau \sim -1.7$ and 
the lower boundary of the flux tube is observed.
For both pixels, the background plasma is not magnetized even 
though we allow the field strength to change in the inversion.
Thus, the flux tube is located in an essentially field-free atmosphere.

The plasma parameters such as the magnetic field strength
significantly change as a function of the optical depth 
in the line forming region due to the presence of the 
flux tube (Fig.~\ref{atmos}). The large gradients are probably 
responsible for the observed asymmetries of the Stokes 
profiles. 
We note that the variation in the plasma 
parameters (as represented by the black areas in Fig.~\ref{atmos}) 
at the same optical depth is relatively small, regardless 
of the considerable differences in the widths and the 
positions of the flux tube, essentially within the line forming 
regions (the region delimited by the two dashed lines in Fig.~\ref{atmos}). 
For pixel\#1, the values above $\log\tau\sim0$ are similar: the broader 
the flux tube becomes, the deeper is located to maintain the similarity 
in the distribution.
Likewise, pixel\#2 has a similar situation. The values of the 
plasma parameters are similar at the same optical depth 
below $\log\tau\sim-2.0$, regardless of the widths and the positions
of the flux tube. Broader flux tubes are positioned higher in the
atmosphere for the same reason.

The rising motion of the flux tube is confirmed by the distributions of
LOS velocity. 
The LOS velocity distributions shown in Figure~\ref{atmos} indicate that 
the flux tube is slightly blueshifted below $\log\tau=-0.7$ for pixel\#1. 
The velocity of the flux tube shows an upward motion with respect 
to the background atmosphere at $\log\tau\sim-1.5$. 
The relative velocity of the flux tube is  $\sim1.3$ km s$^{-1}$. For
pixel\#2, the flux tube is blueshifted, showing LOS velocities up to 1.6 km s$^{-1}$.
The relative velocity with respect to the background is larger than for pixel\#1.

As can be seen in Figure~\ref{atmos}, the inclination angle of the flux tube is
$\sim90^{\circ}$, so that the field lines are perpendicular to the LOS for both pixels. 
The azimuth angle, measured from the east-west direction, is about $160^{\circ}$;
indicating that the field lines are nearly parallel to the line
connecting the negative and positive Stokes $V$ patches of the
second scan. The magnetic field strengths are around 400~G in both cases.

\section{Three dimensional structure and time evolution}
\label{3D}
\begin{figure}
\epsscale{.5} \plotone{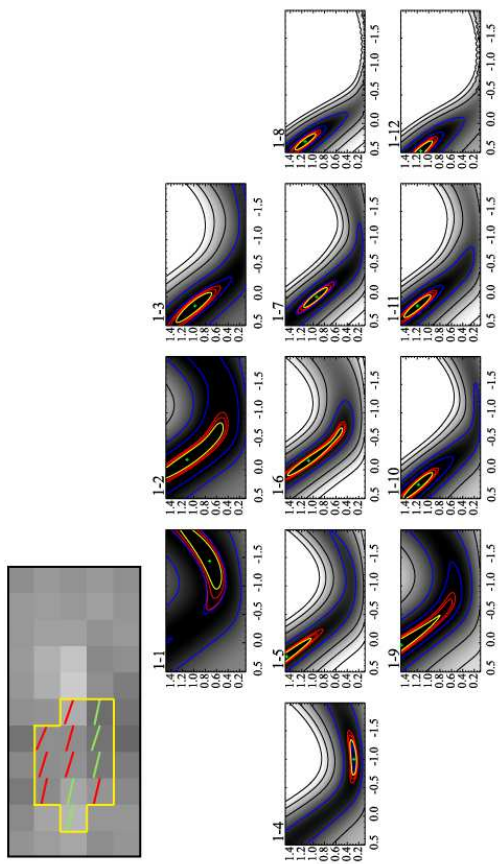}
  \caption{$\chi^{2}$ maps equivalent to those shown in Fig.~\ref{chi} for the first scan ($\Delta t=$0~s).
Pixel 1-6 corresponds to pixel\#1.  
Lower left panel shows the integrated circular polarization signals for these pixels. The green and red lines represent the azimuth direction at $\log\tau=-1$.}
  \label{firstchi}
\end{figure}

\begin{figure}
\epsscale{.7} \plotone{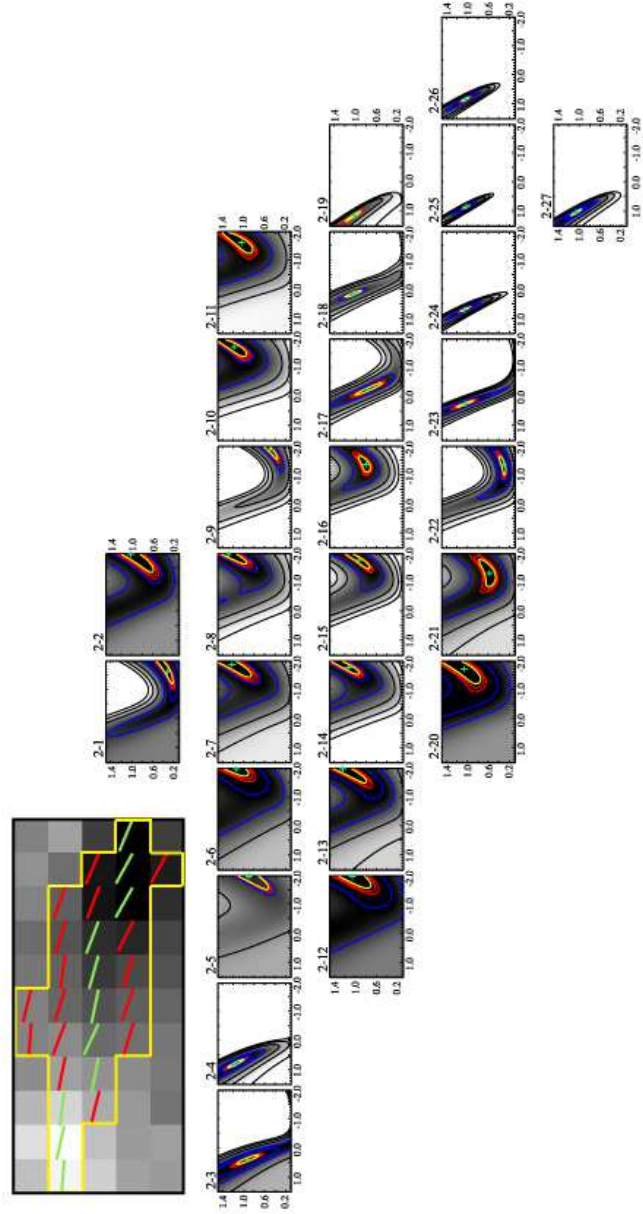}
     \caption
{Same as Fig.~\ref{firstchi}, but for the second scan ($\Delta t=$130~s). Pixel 2-14 corresponds to pixel\#2. }
     \label{secondchi}
\end{figure}

We have analyzed those pixels outlined by red and green contours (See Fig.~\ref{eventfig}),
following the approach we applied to pixel\#1 and pixel\#2 in the previous
section. Figures~\ref{firstchi} and~\ref{secondchi} show the corresponding
$\chi^{2}$ maps for these pixels. Pixel\#1 and pixel\#2 correspond to 
pixel 1-6 in Figure~\ref{firstchi} and pixel 2-14 in Figure~\ref{secondchi}, respectively.

In the first scan, the results are comparable to those found for pixel\#1. The
$1\sigma$ confidence regions are elongated and located low in the
atmosphere. However, this is not the case for pixels 1-1 and 1-4. In
these pixels, although the fits are as good as those in the neighboring pixels, the
location of the flux tube is different.

In the center of the THMF (pixels 1-6 and 1-2), the $1\sigma$ confidence region
is located between $\log\tau=0$ and $\log\tau=-0.7$. There the
position of the flux tube is the highest. In the surrounding pixels (1-3,
1-5, 1-7, 1-9, 1-10, 1-11) the 1$\sigma$ confidence region is located below $\log\tau\sim0$. 
At the one of edges of the THMF (pixels 1-8 and
1-12), the location of the flux tube is even lower. 
In summary, if we go from the center of the THMF to any of its edges, the flux tube
sinks into the photosphere. Notice that for most pixels the lower boundary of the
flux tube is below the continuum formation level.  This results is compatible 
with the presence of  a low laying flux tube with an $\Omega$ shape. Only the
apex of the flux tube is detected in linear polarization (i.e., horizontal fields)
and therefore, the footpoints of the loop are not seen.

In the second scan, in the center of the THMF, the flux tube is located at around
$\log\tau=-1.6$. When we go from the center of the THMF towards any of its footpoints 
(e.g., from pixels 2-15 to 2-19), the flux tube is located deeper in the atmosphere.
Again, this is consistent with a flux tube with a $\Omega$ shape.
The retrieved shape and position of the flux tube from the first and second
scan confirms our observation of a rising $\Omega$~loop.

\section{Properties of transient horizontal magnetic field}
\subsection{Magnetic field topology}
\label{CrossMag}
The lower left part of Figures~\ref{firstchi} and~\ref{secondchi} shows
the azimuth angles of the magnetic field lines retrieved with the SIRGAUS code at
$\log\tau=-1.0$. For both scans, the magnetic field is aligned with the axis of
the flux tube, i.e., with the line connecting the opposite polarity patches. The
averaged azimuth angles around the center of the THMF are 164$^{\circ}$ and
162$^{\circ}$ for the first and second scan, respectively. Since the
azimuth is displayed at the same optical depth for both scans, it means that the direction
of magnetic field is the same in the upper part of the flux tube observed at
$\Delta t=0$~s as in the lower part of the flux tube observed at $\Delta t=$130~s.
This indicates that the analyzed THMF does not show any helical structure,
so that essentially, the flux tube has a toroidal component only.

The pixels in which the azimuthal
directions are shown with green lines in Figures~\ref{firstchi} and~\ref{secondchi}
are selected for plotting the magnetic field strength cross-section of 
the flux tube, along its axis (top panels in Fig.~\ref{crossmap}).
Left and right panels correspond to the first and second scans.
For both scans, the field strength distributions clearly show that the flux tube is located 
deeper in the atmosphere, going from the center to the edge of the THMF.
In the first scan, we detect only the apex of the flux tube, while 
in the second scan the flux tube is already located high in the 
line forming region.
These results clearly indicate a rising $\Omega$-shaped flux tube.
The magnetic field strength of the flux tube is
smaller than 400~G. We do not see any change in magnetic field strength between
the first and second scan.

The white lines plotted over the distributions of the magnetic field strength 
in Figure~\ref{crossmap}
mark its inclination. In the first scan only the horizontal component of
the magnetic field can be detected, $\gamma\sim90^{\circ}$, (there is no Stokes $V$ signal
observed during this scan). The inclination at the central part of the THMF
is also around $90^{\circ}$ in the second scan. However, near the region
where Stokes $V$ signal is detected, the retrieved inclinations
indicate that the field lines are more vertical. The retrieved inclinations nicely delineate
the shape of the flux tube. 

\subsection{LOS velocity topology}
\label{CrossVelo}
\begin{figure}
\epsscale{0.7} \plotone{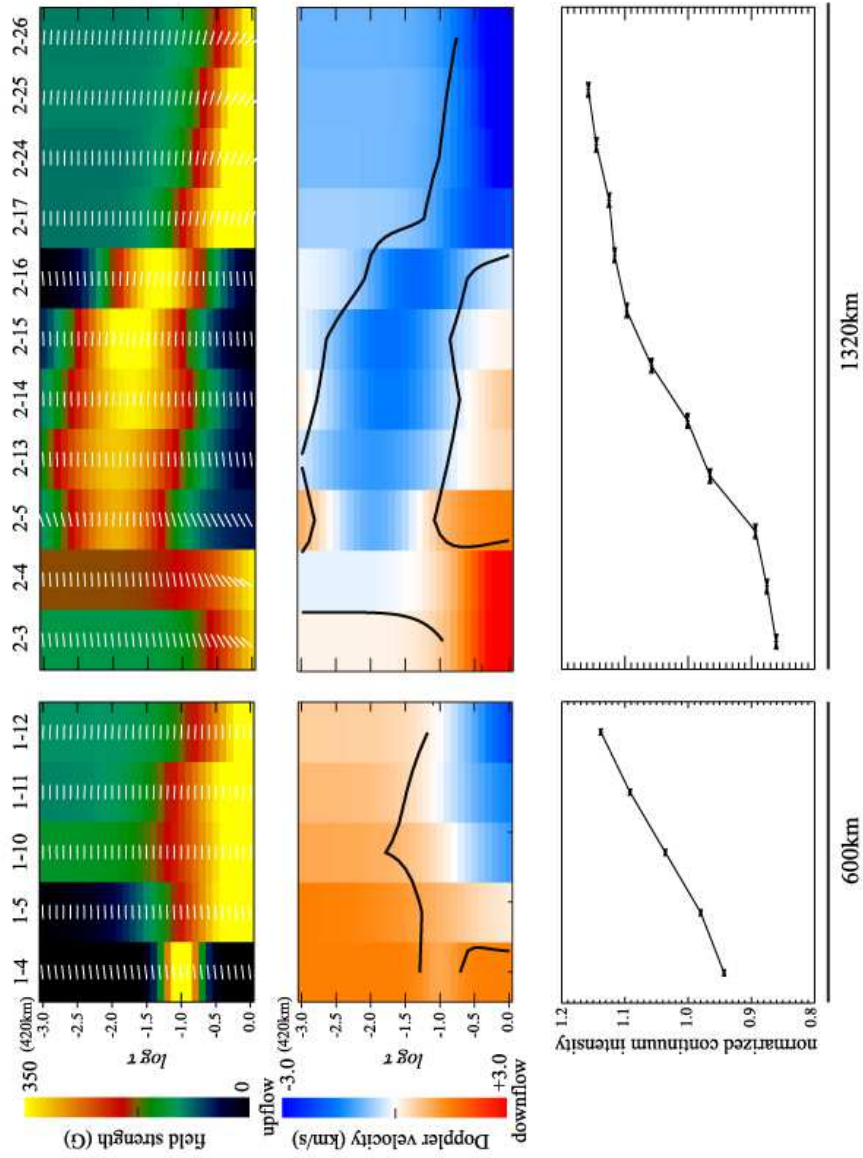} \caption{Top panels: Cross-sectional view of
the magnetic field strength of the THMF for the first ($left$) and the second
($right$) scans.  White lines show the inclination of the magnetic field
vector. Middle panels: Cross sectional view of the Doppler velocity.  The
region with field strength above 150~G is contained with the solid lines.
The shown distributions correspond to the pixels that are indicated by green lines in the azimuth maps from Figs.~\ref{firstchi} and \ref{secondchi}.
For each
pixel, the atmospheric parameters with the minimum $\chi^{2}$ are plotted.
The bottom two panels represent the  normalized continuum intensity. } \label{crossmap}
\end{figure}

The distributions of LOS velocity along the axis of the flux tube can be seen in 
Figure~\ref{crossmap} (middle panels).
Since the plasma velocity depends on 
whether the pixel is located in the granular/intergranular region, the normalized continuum intensities are also displayed in the bottom panels of
Figure~\ref{crossmap}.

Pixels 1-4 and 1-5 have continuum intensities lower
than 1.0, and correspond to the edge of the granule (see also Fig.
\ref{eventfig}). In these pixels, the flux tube show downward motion.
In the same pixel, the background atmosphere shows larger downflows.
From pixel 1-10 to pixel 1-12, the continuum intensity is above 1.0 and becomes larger. 
These pixels correspond to the transition region from the edge of the granular cell to 
the interior of the granule.
There, the flux tube exhibits a strong upward motion, with velocities reaching
2.7~km~s$^{-1}$. In contrast, the background plasma above the flux tube is
slightly redshifted with LOS velocities smaller than 1 km~s$^{-1}$. Note that
the absolute Doppler velocities with respect to the solar surface are shown in
Figure~\ref{crossmap}. \citet{Borrero2002} obtained a two-component model of the
solar photosphere from the inversion of spatially and temporally averaged
solar intensity profiles, and they found upflows of $\sim$1.1~km~s$^{-1}$
in granular cells at around $\log\tau \sim -0.3$ (Note that they only correct 
the gravitational redshift). Thus, the upward velocities of the flux tube we find 
in our analysis are larger than that for typical
granular cells.

In the second scan, the flux tube is located higher in the atmosphere. 
At the apex of the loop (pixel 2-13), the 
continuum intensity is slightly below 1.0. This pixel appears to be located at the
boundary between the granule and the intergranular lane. The pixels around 
the apex of the flux tube show  
upward velocities reaching $\sim3$~km~s$^{-1}$, clearly
indicating the rising motion of the flux tube.

Next, we examine the more vertical portion of the flux tube located in the low
atmosphere. The region of negative polarity (pixels 2-17, 2-24, 2-25, and
2-26) shows the continuum intensities larger than 1.0, corresponding to
the inner part of the granule. There, the flux tube has strong upward velocities ranging from
$\sim2$~km~s$^{-1}$ to $\sim4$~km~s$^{-1}$. These velocity values are larger than that 
of the typical upflows observed in granules.

In the portion of the flux tube with positive polarity (pixels 2-3 and 2-4),  
the continuum intensity is lower than 1.0.
These pixels are located in an intergranular lane. 
The flux tube at these positions has a downward motion ranging from
$\sim2.3$~km~s$^{-1}$ to $\sim4.6$~km~s$^{-1}$ at $-0.5<\log\tau<0$.
Typical downflow velocities in the intergranular lanes in the lower photosphere
($-0.5<\log\tau<0$) are between 2.2 km s$^{-1}$ and 3.7
km s$^{-1}$ \citep{Borrero2002} comparable to what we measured. Therefore, 
the downward motion of the
flux tube may be driven by the convective motion so that 
the flux tube may be forced to be submerged with the downward flow field.

As we have seen, the flux tube is located at around
$\log\tau = 0$ in the first scan ($\Delta t=0$~s), 
while in the second scan ($\Delta t=130$~s), the
center of the flux tube is located at around $\log\tau=-1.7$. This clearly
indicates that the flux tube goes upward and reaches higher photospheric layers with time. 
The geometrical distance between $\log\tau=0$ and $\log\tau=-1.7$ is around 244~km.
Given the time difference of 130 s between these two scans, the rise velocity
of the flux tube is 1.9~km~s$^{-1}$. This is consistent with the inferred upward motion
of 2~km~s$^{-1}$ for the flux tube (See middle panels of
Fig. \ref{crossmap}).

\subsection{Cross sectional shape of flux tube}
\label{CrossShape}
To estimate the width of the flux tube, we need to find a pixel, where the $1\sigma$
confidence regions in the $\chi^{2}$ maps (marked by the yellow contour in Figures~\ref{firstchi}
and~\ref{secondchi}) are not elongated and where we observe both the upper and
lower boundary of the flux tube. In the second scan, the center of the analyzed THMF is already too
high for our purpose. We select pixel 2-16. This pixel is located at the
center of the THMF, and its $\chi^{2}$ 1$\sigma$ confidence levels are not spanning over a 
wide range of $G_{width}$ and $G_{pos}$. The upper and the lower boundary are
within the line forming region, as well (See the top right panel from Fig.~\ref{crossmap}).

For pixel~2-16 the position of the flux tube is located at $\log\tau=-1.4$ 
and the width corresponds to 0.8 in $\log\tau$ units as given by the $\chi^{2}_{min}$. 
We define the radius of the flux tube along the LOS as the FWHM of
the Gaussian function that represents it. In this case the vertical extension of the
flux tube is around 190~km. We estimate the horizontal size to be
approximately 360~km, since the THMF occupies three pixels (2-9, 2-16 and 2-22) 
at around the location of pixel~2-16. It should be kept in mind that the real horizontal size
might be larger since a weaker Stokes $Q$ and $U$ signal can be observed surrounding the 
THMF. The discrepancy between the vertical and horizontal size of the flux
tube indicates that the flux tube is squeezed in the vertical direction.

\subsection{Total magnetic flux carried by THMF}
\label{Magflux}
\begin{figure}
\epsscale{0.9} \plotone{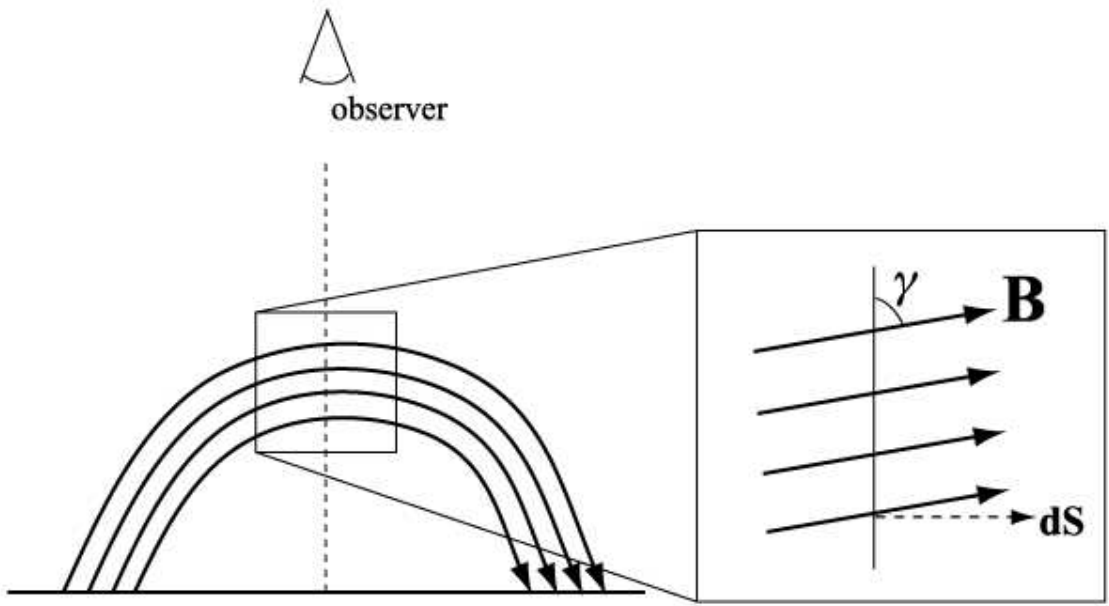} \caption{Representation of the vector field
and the surface used for the definition of the horizontal magnetic flux.} \label{flux}
\end{figure}

In this section we estimate the magnetic flux $\Phi_{H}$ carried by the THMF, since we know the vertical (along the LOS) and lateral cross section of the flux tubes as shown in the previous sections.
Let $\vec{B}(z)$ be the magnetic field vector with inclination $\gamma(z)$,
$z$ being the geometrical height, and the SOT/SP pixel size $l'$ of about 120~km. 
The per-pixel magnetic flux $\varphi_{h}$ through the surface $\vec{S}$ 
(see Fig.~\ref{flux}) is given by
\begin{align}
\varphi_{h}& = \int_{S}^{} \vec{B} \cdot  d\vec{S} \notag\\
& \approx \int_{z_{0}(\log\tau=0)}^{z_{1}(\log\tau=-2.5)} B(z) \cos (90^{\circ}-\gamma(z)) (l' dz). 
\label{eq:flux1}
\end{align}
The integration range is set to cover the line forming region of the \ion{Fe}{1} lines. 
To estimate  the total magnetic flux $\Phi_{H}$ carried by the THMF
we sum up the $\varphi_{h}$ values for pixels of 2-1, 2-7, 2-14, and 2-20 from the second scan.
The total magnetic flux $\Phi_{H}$ turns out to be $3.1\times10^{17}$ Mx. This value
is in agreement with that derived by \citet{Jin2009}, who assumed that
the vertical extent of the flux tubes is the same as that of the scale height of 100
km. In our analysis we resolve the vertical structure of the magnetic flux tube using SIRGAUS.
This has allowed us to derive the magnetic flux more directly.
Note that the total magnetic flux may be different depending on the polarization threshold 
used in the analysis, so that our magnetic flux estimate represents a lower limit.

\section{Summary and discussions}

\subsection{3D magnetic structure and evolution}
We have analyzed a magnetic phenomena in which 
horizontal magnetic field patch appears first at the edge of a granular cell at
$\Delta t=0$ s. Later at $\Delta t=130$ s, the size of the patch harbouring horizontal
fields becomes larger in size, and vertical magnetic fields with opposite polarity appear
at both ends of the magnetic patch. Then (at $\Delta t=260$ s), the horizontal
magnetic field disappears, while the vertical magnetic fields remain and separate. The
whole process takes about 4 min. This event corresponds to a typical example of the
so called transient horizontal magnetic field (THMF).

We have successfully applied the SIRGAUS inversion code to this event and analyzed
its 3D magnetic structure. The results show that the flux tube occupies only a fraction 
of the line formation region of the  \ion{Fe}{1} lines at 630.2~nm and that it has an upward movement.
The magnetic flux tube rises from $\log\tau \sim 0$ (0~km) to $\log\tau \sim -1.7$
(244~km) between the first and the second scan that are separated in time by
130~s. The Doppler velocity of the flux tube consistently indicates the upward
motion.
The magnetic field strength of the flux tube is about 400~G.
The azimuth direction of the magnetic field is aligned with the axis of the flux
tube, and the flux tube does not show any helical structure.

So far, to determine whether the observed magnetic loops correspond to the rising
$\Omega$ loops or to submerging U loops, 
Doppler velocity measurement have been used \citep[e.g.,][]{MartinezGonzalez2009}.  
Our analysis with SIRGAUS presented here shows that the flux tube with shallow
but clear $\Omega$ shape rises. 
We observe only a small part of the shallow $\Omega$ loop above the continuum formation 
layer and this might be just a part of a larger-scale magnetic structure located in the bulk of
the convection zone. 

\subsection{Force balance of flux tube}
\label{forcebalance}
The field strength of the flux tube is
$\sim$400 G in the first scan ($\Delta t=0$~s). Taking the velocity
of the convective flow to be $v=2\times10^{5}$~cm~s$^{-1}$ and the mean density
at the base of the photosphere ($\log\tau\sim0$) to be $\rho=$2.7$\times$
10$^{-7}$~g~cm$^{-3}$, we find that the equipartition field strength is
$B_{eq}=(4\pi \rho)^{1/2}v=370$~G. Therefore, the inferred
field strengths are comparable to $B_{eq}$. This is
consistent with the fact that the flux tube seems to maintain its
integrity during its rise from the lower photosphere upwards.

In this section, we compare the buoyancy force, $F_{b}=(\rho_{e}-\rho_{i})g$, with the
magnetic tension force, $F_{T}=B^{2}/4\pi L$, acting on the flux tube at $\Delta
t=$130~s. Here $\rho_{i}$ and $\rho_{e}$ are the plasma density inside and
outside the flux tube respectively, $g$ is the gravity acceleration of
$2.7\times10^{4}$~cm~s$^{-2}$, and $L$ is the curvature radius of the $\Omega$
loop. 
Here we use all the plasma parameters (temperature, field strength, and the density) at the
height where the flux tube is located and the 3D geometrical scale of the flux tube (the length 
and the cross section). 
From the separation of the footpoints in the second scan (which is about 1320~km) and
from the averaged magnetic field inclination in the footpoints
(about 47$^\circ$), we estimate $L$ to be 1250~km. The comparison is made at the apex 
of the flux tube which is located above $\log\tau\sim-1.0$.
There the temperature $T$ is $\sim$4950~K (corresponding to the averaged value 
from $\log\tau=-1.0$ to $-$2.0 at pixel 2-14). Assuming that the temperature is the same 
inside and outside of the flux tube, $F_{b}$ can be written as $F_{b}=\frac{B^{2}}{8\pi}
\cdot \frac{m_{p}g}{2k_{B}T}$, where $m_{p}$ is the proton mass, and $k_{B}$
the Boltzmann constant. The ratio between $F_{b}$ and $F_{T}$ is then
$\frac{F_{b}}{F_{T}}=\frac{m_{p}gL}{4k_{B}T}\sim2$. This clearly indicates that
the buoyancy force, $F_{b}$, participates in the emergence process of the magnetic flux
tube.

If we suppose that the stationary state is reached, then the buoyancy force,
the magnetic tension force, and the resultant drag force associated with the rising
motion are balanced:
\begin{align}
(\rho_{e}-\rho_{i})g \cdot al-\frac{B^{2}}{4\pi L} \cdot
al-0.5C_{d}\rho_{e}u^{2}l & = \frac{B^{2}}{8\pi} \cdot
\frac{m_{p}g}{2k_{B}T}\cdot al-\frac{B^{2}}{4\pi L} \cdot
al-0.5C_{d}\rho_{e}u^{2}l \notag\\ & = 0.
\label{eq:force}
\end{align}
With $C_{d}$ being the aerodynamic coefficient of the order of unity
\citep{parker1979}, and $u$ the relative velocity of the flux tube with respect
to the velocity of the surrounding plasma. 
$a$ and $l$ are the vertical extent and the horizontal size of the flux tube, respectively. 
We take $\rho_{e}$ to be $7.78\times10^{-8}$~g~cm$^{-3}$ (corresponding to the mean
density from $\log\tau=-1.0$ to $-2.0$ at pixel 2-14). The relative velocity $u$ obtained from Eqn.~(\ref{eq:force}) is then 2.3~km~s$^{-1}$, which is consistent with the observed relative Doppler velocity associated to the flux tube.

As shown in section~\ref{CrossShape}, the vertical extent of the flux tube ($a=190$ km) 
is smaller than its lateral size ($l=360$~km). Therefore, the cross section of the flux tube
has a flattened shape. \citet{magara2001} studied the emergence of large flux tubes with larger 
field strengths which form active regions. He found that the top of the loop decelerates when it 
reaches 
photospheric layers while the bottom of the flux tube continues to rise. 
The reason is that the photosphere is a convectively stable layer. Therefore the cross
section of the tube changes from the circular shape to horizontally extended shape.
The same process may take place for granular-sized magnetic flux tubes.
\citet{Steiner2008} pointed out that an emerging flux tube can get pushed to the middle and 
upper photosphere by overshooting convection, forming atmospheric layers full of horizontal
fields. This may also contribute to the flattening of the flux tube.

\subsection{Magnetic flux and energy}
The horizontal magnetic flux $\Phi_{H}$ is estimated to be $3.1\times10^{17}$
Mx near the apex of the flux tube. Recently \citet{Ishikawa2009} reported that the occurrence
rate of THMFs is about $1.1\times10^{-10}$ km$^{-2}$ s$^{-1}$ in a quiet Sun region.
Also, \citet{Tsuneta2008ApJ} and \citet{Itoh2009} have shown the presence of horizontal
magnetic fields ubiquitously in the polar region, and that there is no difference on
the intrinsic magnetic field strengths between the quiet Sun and the polar region. \citet{Lites2009} suggested that
the uniform fluctuations of longitudinal magnetogram signals over the whole disk
observed by \citet{Harvey2007} are the same phenomenon as the horizontal fields
observed by \emph{Hinode}. All these results together indicate that THMFs would have the
same occurrence rate all over the solar surface. 
Combining these observations, we conclude that THMFs
could carry a total magnetic flux of $1.8\times10^{25}$~Mx per day
to the higher atmosphere what corresponds to a total of $7\times10^{28}$~Mx in one solar
cycle (11 year). This estimation is based on the assumption that all THMFs reach higher 
atmospheric layers and that all of them carry the same amount of magnetic flux $\Phi_{H}$ 
(given in this paper). 
Although our calculations are based only upon a single event, similar values  have been 
found also by  \citet{Jin2009} and \citet{MartinezGonzalez2009}.
The magnetic flux carried by THMFs is much larger than the total
magnetic flux in sunspot regions during one solar cycle which corresponds to about $10^{25}$ Mx \citep{Harvey1993}.

\citet{Ishikawa2009} have estimated the magnetic energy flux carried by THMFs, and
found that the amount of energy flux is comparable to the total chromospheric and coronal
energy loss, assuming that all THMFs reach above the photosphere. In order
to perform a more precise estimation of the total magnetic flux and the total magnetic
energy that THMFs could carry to higher atmospheric layers, we need to know what
percentage of THMFs reach the chromosphere or higher layers in addition to the precise
estimation of the magnetic flux of each THMF. 
Such statistical analysis was done by \citet{MartinezGonzalez2009}, who
pointed out that 23\% of the horizontal fields that emerge with clear bipolar footpoints in
the photosphere reach the low chromosphere. 

\subsection{Disappearance of THMFs}
We show that the THMF analyzed in this paper is rising through the line forming
layer of the \ion{Fe}{1} 630.2~nm lines and that it reaches higher atmospheric layers. 
The flux tube essentially maintains its integrity during the rising motion. At
the beginning of this paper, we addressed three different possibilities of the
THMFs disappearance. The observed event is clearly into the third of the suggested
mechanisms, i.e., the THMF goes through the photosphere reaching the layers above the line 
forming region of  \ion{Fe}{1} 630.2~nm lines.
Some THMFs do not show any conspicuous footpoints in their lifetime
\citep{Ishikawa2008, Jin2009}. Such THMFs possibly harbor weaker magnetic
fields than the equipartition field strength. Thus, the convective motion may
destroy the flux tube before it reaches the upper photosphere, or it may force
it to submerge below the photosphere. 

\acknowledgements The authors want to express their sincere appreciation to Bellot Rubio for 
his pioneering work, and thank him for allowing us to use the SIRGAUS software.  The
authors are thankful for the encouragement by Bellot Rubio, Ruiz Cobo, Del Toro
Iniesta, and Orozco Su{\'a}rez. 

\emph{Hinode} is a Japanese mission developed and launched by ISAS/JAXA, with
NAOJ as a domestic partner and NASA and STFC (UK) as international partners. It
is operated by these agencies in co-operation with ESA and NSC (Norway). 
The financial support from GA~AS~CR IAA~300030808 to Jan Jur\v{c}\'{a}k is gratefully
acknowledged.

\end{document}